\documentclass[article,showpacs,preprintnumbers,amsmath,amssymb,usletter]{revtex4}

\usepackage{amsmath,amssymb,bbm,epsfig}

\newcommand{\be}{\begin{equation}}
\newcommand{\ee}{\end{equation}}
\newcommand{\ba}{\begin{array}}
\newcommand{\ea}{\end{array}}
\newcommand{\bqa}{\begin{eqnarray}}
\newcommand{\eqa}{\end{eqnarray}}

\newcommand{\um}{\mathbbm{1}}

\newcommand{\ket}[1]{\ensuremath{| #1 \rangle}}
\newcommand{\prj}[1]{\ensuremath{| #1 \rangle \langle #1 |}}

\begin{document}

\title{Optimal correction of independent and correlated errors}

\author{Sol H. Jacobsen\footnote{sol.jacobsen@utas.edu.au} and Florian Mintert}

\affiliation{Freiburg Insitute for Advanced Studies, Albert-Ludwigs University of Freiburg, Albertstr. 19, 79104 Freiburg, Germany}

\date{\today}

\begin{abstract}
We identify optimal quantum error correction codes for situations that do not admit perfect correction. We provide analytic $n$-qubit results for standard cases with correlated errors on multiple qubits and demonstrate significant improvements to the fidelity bounds and optimal entanglement decay profiles.
\end{abstract}


\maketitle

\section{Introduction}
\label{sec:Introduction}
The goal of quantum error correction (QEC) is to transmit quantum information reliably despite noise and decoherence resulting from unavoidable environment coupling. On the one hand, QEC is essential for the practical implementation of quantum communication protocols and quantum computational algorithms, but, on the other hand, QEC also provides an interesting forum for examining the conditions in which generally irreversible open system dynamics can be undone.

Reversibility can only be achieved for specific decoherence models, and real open system dynamics will typically always deviate from such idealized scenarios. In practice one may achieve reversibility for the predominant decoherence effects, but typically not for the remaining subordinate processes. Consequently, if there is a clear separation in probability regimes for the various detrimental processes, {\it i.e.} errors that may occur, the best strategy is likely to target perfect correction of the more probable errors at the expense of leaving less likely errors uncorrected. This is the strategy currently employed by most QEC protocols. However, there is a plethora of common scenarios in which the simplification to separable probability regimes is insufficient, e.g. if, in addition to single qubit errors, there are correlated errors caused by fluctuations in global control fields, such as the laser that creates an optical lattice for neutral atoms \cite{BlochReview} or a field gradient that induces an interaction between $N$ ions \cite{MintertWunderlich}. Several important steps have already been taken in terms of quantifying the impact of correlated noise and calculating accuracy thresholds for concatenated codes and fault tolerant computation (see for example 
$[3\!-\!7]$), but the simplification remains a persistent problem. In the current paper we turn instead to the idea of \textit{approximate} QEC: if a clear separation of probability regimes is not evident, it might instead be better to abandon the goal of perfect correction of the high-probability subset of errors \cite{Leungetal1997}. In this paper, we will present a method to determine optimal QEC protocols for the cases where a typical separation of probability regimes with dominant single qubit errors is not present.

Quantum dynamics can be described in terms of quantum channels $\Omega$, {\it i.e.} completely positive maps
\begin{eqnarray}
\varrho(t)=\Omega(\varrho(0))=\sum_{i=1}^{N}E_i(t) \varrho(0) E_i^\dagger(t)\ ,\label{Eqn:genchannel}
\end{eqnarray}
that connect a system's density matrix $\varrho$ at an initial time $t=0$ and a later time $t$.
The specificities of the actual dynamics are characterized by the $N$ time-dependent Kraus operators $E_i(t)$.
If the Kraus operators satisfy the condition $\sum_{i=1}^NE_i^\dagger(t) E_i(t)=\um$, then the channel is trace preserving and thus deterministic.
Conversely, if $\sum_{i=1}^NE_i^\dagger(t) E_i(t)<\um$ then the channel is probabilistic and there is a finite probability for a process that is not described by the channel to occur.

In the context of QEC, a deterministic channel is typically divided into a correctable and an uncorrectable part:
\begin{eqnarray}
\Omega_t(\varrho)=\Omega_c(\varrho)+\Omega_u(\varrho)\ ,
\end{eqnarray}
such that $\Omega_c(\varrho)$ has an inverse channel for initial states in a well-defined subspace of the full system's Hilbert space.
This subspace is called the code and $\Omega_c$ is correctable for initial states in the code. If the contributions from $\Omega_u$ vanish for $t=0$ and result in deviation from trace conservation of the correctable channel $\Omega_c$ which grows slower than linearly with $t$, then arbitrarily good correction of $\Omega_t$ can be achieved \cite{FTQEC}.

The question of which states render a given channel correctable has a surprisingly simple answer, encapsulated by the necessary and sufficient conditions for the existence of a recovery operation \cite{KnillLaflamme, NielsenChuang}. That is, if and only if there is a projector $P$, {\it i.e.} code, such that the Kraus operators describing the channel $\Omega$ satisfy the relation
\begin{eqnarray}
PE_i^\dagger E_j P=\alpha_{ij}P,\label{PEEP}
\end{eqnarray}
then there is a deterministic quantum channel $\Omega^{-1}$, such that $\Omega^{-1}(\Omega(\varrho))=p\varrho$ with $p=\mbox{Tr}\Omega(\varrho)$ for all states that satisfy $P\varrho P=\varrho$, \textit{i.e.} all states in the space defined by $P$. Here $\alpha$ is a Hermitian matrix with complex elements and the time dependence of the Kraus operators $\left\{E_i(t)\right\}$ is left implicit. 

In the standard implementation of QEC, the conditions ~(\ref{PEEP}) are satisfied for the Kraus operators of $\Omega_c$ only, while those of $\Omega_u$ reflect the finite (and presumed negligible) probability for an uncorrectable error. In the following we consider the scenario that this probability is not negligible, and examine the correctability of the full trace preserving channel. This more general case occurs when, for example, the probability for an error associated with the uncorrectable channel grows linearly or faster in $t$, or sufficiently fast repetition of $\Omega_c^{-1}$ is not possible. 

The task at hand is thus the rigorous identification of optimal codes for conditions in which perfect correction of the errors is not possible -- the regime of approximate quantum error correction (AQEC). Such endeavours have been attempted previously, and it has been confirmed that the performance of quantum codes through the full noisy channel can be improved through the relaxation of the conditions (\ref{PEEP}) \cite{Leungetal1997,NgMandayam, MandayamNg}. A typical measure of the performance of a given code is the fidelity between the input and output states after noise, recovery and decoding. Finding truly optimal codes in terms of the fidelity would thus typically require numerical optimisation over all possible encodings and recovery maps (see e.g. \cite{NgMandayam}), which is extremely computationally demanding if not practically impossible unless one fixes one or separates the two optimisations. For this reason a number of approaches have been found to establish (near-optimal) bounds on the fidelity of codes given the noisy channel (see for example 
$[12,14\!-\!17]$). However, recent work has indicated that targeting a state's remaining entanglement directly is of interest and can lead to new analytic insight $[18\!-\!20]$,
since its decay is explicitly connected to reversibility \cite{SW2002, HorodeckiReview, Schumacher1996}. In what follows, we will present a straightforward method for finding optimal quantum codes given a noisy channel, which can be assessed without the need to consider recovery directly, and we will demonstrate significant improvements in code performance for some standard examples using both fidelity bounds and entanglement decay.

The rest of the paper is organised as follows: In Section \ref{sec:2} we introduce the nomenclature we will use in determining optimal codes in the AQEC regime before outlining the method and example. In Section \ref{sec:Opt} we provide detailed analysis of the procedure for a $3$-qubit example before providing the $n$-qubit generalisation with analytic results. We conclude in Section \ref{sec:Discussion} with some discussion and prospects for further work.

\section{Finding optimal AQEC codes}
\label{sec:2}
As we saw in the Introduction, it will be useful to define both fidelity bounds and the entanglement dynamics associated with quantum error correcting codes in order to construct optimal procedures for information protection. Here we begin by describing these procedures.

\subsection{Fidelity bounds}
We saw that the possibility of recovering the original message by employing a particular quantum code is encapsulated in criterion~(\ref{PEEP}), and we introduced the scenarios in which no projectors can be found to satisfy those conditions. Perfect correction in the sense of (\ref{PEEP}) would correspond to a fidelity $F=1$ between the input state and output state after noise, recovery and decoding. For AQEC, $\epsilon$-correctable codes are those which result in a fidelity of at least $\sqrt{1-\epsilon}$. The fidelity loss $\eta$ for a given code $\mathcal{C}$ and recovery channel $\mathcal{R}$ is defined as:
\begin{eqnarray}
\eta\equiv 1-\mathop{\mbox{min}}_{|\psi\rangle\in \mathcal{C}} F^2(|\psi\rangle, \mathcal{R}\cdot \Omega).
\end{eqnarray}
The optimal fidelity loss $\eta_{op}$ would thus be the minimum of $\eta$ over all possible recovery channels $\mathcal{R}$, and the code would be $\epsilon$-correctable if it has $\eta_{op}\leq \epsilon$.

To determine the bound on the fidelity between input and output states, we will consider the AQEC conditions as presented in \cite{NgMandayam}:
\begin{eqnarray}
PE_i^\dagger \Omega(P)^{-1/2}E_jP=\beta_{ij}P+\Delta_{ij},\label{AQECNg}
\end{eqnarray}
where $\beta\equiv\sqrt{\alpha}$ and $\Omega\sim\{E_i\}$ is the noisy channel. The fidelity loss can then be written:
\begin{eqnarray}
\eta=\mathop{\mbox{max}}_{|\psi\rangle \in \mathcal{C}} \sum_{ij}[\langle\psi|\Delta_{ij}^\dagger\Delta_{ij}|\psi\rangle-|\langle\psi|\Delta_{ij}|\psi\rangle|^2],\label{eta}
\end{eqnarray}
such that $\mathcal{C}$ is $\epsilon$-correctable if $\eta\leq \epsilon$, and can be evaluated without requiring knowledge of the optimal recovery. That is, (\ref{eta}) provides a guarantee on the maximum fidelity loss for given noise.

\subsection{Entanglement dynamics}
Great advances in the development of near-optimal recovery procedures and analytic bounds on the code performance have been achieved by considering a set of orthogonal states $\ket{\Psi_i}$ that span the code, \textit{i.e.} $\sum_{i=1}^d\prj{\Psi_i}=P$, and the corresponding entangled state 
\begin{eqnarray}
\ket{\kappa}=\sum_{i=1}^d\frac{1}{\sqrt{d}}(\ket{\phi_i}\otimes\ket{\Psi_i})\label{Kappa}
\end{eqnarray}
between an ancillary system with orthonormal states $\ket{\phi_i}$ and the system of interest (see for example [14-17]). 
The system is then affected by the noisy channel with no dynamics in the ancillary system. In fact, we shall go on to show that entanglement preservation of such states corresponds directly with the choice of codes, and that optimal code choice will lead to maximal entanglement preservation for such states. 

If one has a perfect code, this implies that the state of the full composite system can be recovered after the effect of the noisy channel through a recovery operation that is acting on the system of interest only, but not on the ancillary system. In terms of entanglement theory, this means that the original state can be recovered through local operations only. Since the initial state and the final state (after the noise and recovery) have the same entanglement content (it is the same state), and since entanglement cannot increase through local operations, this implies that entanglement did not decrease through the effect of the noisy channel on the perfect code. For imperfect codes therefore, the loss of entanglement, via an appropriate entanglement monotone, is a natural tool to characterise the correctability of a noisy channel.

The entanglement dynamics of the density matrix $\rho=|\kappa\rangle\langle\kappa|$ corresponding to (\ref{Kappa}) as the coded qubit is transmitted through the noisy channel are calculated using the Lindblad master equation (see for example \cite{NielsenChuang}):
\begin{eqnarray}
\frac{d\rho}{dt}=\sum_j\left[2L_j\rho L_j^\dagger-\left\{ L_j^\dagger L_j,\rho\right\}\right].\label{Lindblad}
\end{eqnarray}
Here $L_j$ are the Lindblad operators, which correspond to the types of error expected for the channel, and may be arbitrary errors in general. For pure dephasing errors as in Table \ref{zchannel} in Appendix \ref{App:tPlots}, the Lindblad operators take the associated form $L_1=\sqrt{\gamma_1}\sigma_z\otimes I \otimes I$ etc., where $\gamma_1$ is the Lindblad rate for errors on single qubits, $\gamma_2$ the rate for errors on two qubits and so on.
In turn, the entanglement dynamics can be represented using an appropriate measure and in our example we use Negativity \cite{VidalWerner}:
\begin{eqnarray}
N(\rho)=||\rho^{T_b}||-1=\sum_i||\lambda_i||-\lambda_i.\label{Eqn:neg}
\end{eqnarray}
Here $||...||$ denotes the trace norm and $\lambda_i$ are the eigenvalues of $\rho^{T_b}$ which denotes the partial transpose of $\rho$ over subsystem $b$.

\subsection{Method}
Evaluating (\ref{eta}) provides us with a way of comparing the behaviour of particular codes and channels but in practice the search for truly optimal codes for given noise is still an extremely computationally demanding numerical task. For our approach to finding optimal codes we shall thus begin with a simpler function characterising deviation from (\ref{PEEP}) and show that this method leads to improvements in both the fidelity bounds and entanglement decay rate, with new analytic insight of the code performance.

Consider the function $\delta_c$ characterising the magnitude of deviation from the conditions of perfect recoverability (\ref{PEEP}):
\bqa
\delta_c&=&\sum_{i,j}\mbox{Tr}(\Lambda_{ij}\Lambda_{ij}^\dagger),\hspace{0.5cm}\mbox{ with}\label{fn}\\
\Lambda_{ij}&=&PE_i^\dagger E_j P-\alpha_{ij}P,\hspace{0.2cm}\mbox{ and}\label{PEEPe}\\
\alpha_{ij}&=&\frac{\mbox{Tr}\ (PE_i^\dagger E_j P)}{\mbox{Tr} (P)}\ .
\eqa
Here the error set $\left\{E_i\right\}$ is now extended to include the operators of $\Omega_u$, thus comprising the full trace-preserving channel. If it were possible to correct this extended set fully, then $\Lambda_{ij}\rightarrow 0$ and (\ref{PEEPe}) would reduce to (\ref{PEEP}). Thus as an initial hypothesis the condition that $\delta_c$ be minimal for a good code seems plausible. The notation (\ref{PEEPe}) for deviation from (\ref{PEEP}) was also discussed in \cite{BenyOreshkov2010} and \cite{NgMandayam, MandayamNg}, in which general necessary and sufficient conditions for approximate operator QEC codes were discussed, along with a suggested relationship between the deviation parameter $\Lambda_{ij}$ and the optimal fidelity loss. In section \ref{sec:entanglement} we demonstrate direct correlation between the magnitude of deviation (\ref{fn}) and the rate of entanglement decay, giving a clear physical interpretation to the otherwise quite abstract $\Lambda_{ij}$.

For illustrative purposes we shall follow throughout the explicit example of a complete dephasing channel, through which one typically protects the message -- a unit of information; a quantum bit -- by encoding the message into a state with more qubits. The simplest examples of such encodings are the archetypal three-qubit repetition codes \cite{Shor1995, NielsenChuang} which, despite their simplicity, remain important test cases for the theoretical development and experimental implementation of quantum computing \cite{Reedetal2012}. The standard three-qubit code for a dephasing channel is
\begin{eqnarray}
P_3=|+++\rangle\langle+++|+|---\rangle\langle---|,\label{dephcode}
\end{eqnarray}
with
\begin{eqnarray}
|\pm\rangle=\frac{1}{\sqrt{2}}(|0\rangle\pm|1\rangle).
\end{eqnarray}
Error detection with such codes uses the majority-rule principle, which fails to produce the correct result if in fact two or more errors occurred. Essential to the correction procedure for repetition codes is thus the restrictive assumption that the channel imparts only single, uncorrelated errors on the coded states, an assumption that we remove in the following.

Upon transmission of the message, the noisy channel acts on the information according to equation (\ref{Eqn:genchannel}). Errors on qubits are effected by the generators of $SU(2)$ (i.e. the Pauli matrices) supplemented with the identity operator, and result in bitflip or dephasing errors or a combidation of the two. For pure dephasing our building blocks consist of the identity $I$ and dephasing error $\sigma_z$:
\begin{eqnarray}
I=\left(\!\begin{array}{cc} 1&0\\0&1\end{array}\!\right),\hspace{0.2cm} \sigma_z=\left(\!\begin{array}{cc} 1&0\\0&-1\end{array}\!\right).\label{errors}
\end{eqnarray}
The code space is a tensor product of the individual qubit Hilbert spaces, and so the operation elements are constructed accordingly, with the full three-qubit dephasing channel given explicitly in Table \ref{zchannel} in Appendix \ref{App:tPlots}. We assume the probability for single and two-qubit errors are independent of the choice of qubits, and since the channel is trace preserving $p_0, p_1, p_2$ and $p_3$ respectively give the probabilities of errors occurring on zero, one, two or three qubits simultaneously. Note that to take account of all permutations, the full probability of a single error occurring on any one of the three qubits is thus $3p_1$ (or $3p_2$ for errors on two of the three qubits). The procedure for constructing the channel for other types and combinations of errors and codes follows intuitively.

\section{Code optimisation}
\label{sec:Opt}
We have begun with the hypothesis that minimising $\delta_c$, \textit{i.e.} the magnitude of deviation from (\ref{PEEP}), will produce optimal codes, and turn now to the question of identifying such codes. For this purpose, new codes $P'_n$ are constructed via random unitary transformations of the original code $P_n$:
\begin{eqnarray}
P'_n=UP_nU^\dagger.\label{P'}
\end{eqnarray}
The transformations are applied prior to the noisy channel and the parameters of $U$ are kept entirely general without imposing the requirement that the code should also completely correct single errors, potentially allowing for a tradeoff for better performance with the full channel. The transformation $U$ is then optimised so as to minimise $\delta_c$. The results of this procedure lead, quite unexpectedly, to analytic expressions for regimes of optimal codes, and we will first present the details for the explicit $3$-qubit dephasing example before describing the generalisation to $n$-qubits.

\subsection{Optimal $3$-qubit codes for known channels}\label{Sec:3qubit}
While the dephasing code (\ref{dephcode}) is commonly referred to as the code to solve errors on single qubits, such a code also satisfies (\ref{PEEP}) when errors occur in other combinations from the full channel (e.g. errors occur on zero or two qubits only) since the code cannot distinguish between these pairs of Kraus operators. That is, the code cannot distinguish between $E_0$ and $E_7$, $E_1$ and $E_6$, $E_2$ and $E_5$, and $E_3$ and $E_4$ of Table \ref{zchannel} in Appendix \ref{App:tPlots}. The probabilities corresponding to these combinations result in $\delta_c=\frac14 p_0 p_3+ \frac34 p_1 p_2$. This is thus a target that must be beaten for any new code to improve on the results of the standard code.

An exhaustive search of the parameter space of the above model reveals that optimal performance for dephasing channels contains a clear delineation between two regimes of optimal codes. The first is the known regime in which (\ref{dephcode}) is optimal, and the second is given by \footnote{Note that the other two permutations of these alternative codes also produce optimal behaviour: they contain different violating Kraus combinations but the same corresponding probability combinations.}
\begin{eqnarray}
P'_3=|0++\rangle\langle 0++|+|1--\rangle\langle 1--|\ .
\label{newcode1}
\end{eqnarray}
Interestingly, qubit rotations such as this were also found to improve robustness for the noisy evolution of graph states in another context \cite{AolitaAcin2010}. For this new optimal code, the Kraus combinations that contribute to violation of (\ref{PEEP}) for the full channel are the following elements in $\Lambda_{ij}$: $E_0^\dagger E_1$, $E_2^\dagger E_4$, $E_3^\dagger E_5$ and $E_6^\dagger E_7$. Since this means there are no violating pairs with probabilities $p_0p_2$ or $p_0p_3$, such a code satisfies (\ref{PEEP}) completely for zero or double, and separately zero and triple errors (e.g. if the channel contains no errors on single qubits). By substituting the corresponding probabilities from Table \ref{zchannel} we find $\delta_c=\frac14 p_0 p_1+ \frac12 p_1 p_2+ \frac14 p_2 p_3$ for the new code. By comparing expressions for the contributions to $\Lambda_{ij}$ in (\ref{PEEPe}) for the two codes, we can thus construct an inequality which dictates their regimes of optimality:
\begin{eqnarray}\label{3codeIneq}
2(\sqrt{p_0 p_3}+3\sqrt{p_1 p_2})\!>\!2(\sqrt{p_0 p_1}+2\sqrt{p_1 p_2}+\sqrt{p_2 p_3}).\hspace{0.3cm}\label{3ineq}
\end{eqnarray}

\begin{figure}[h!]
\centering
\includegraphics[width=10cm]{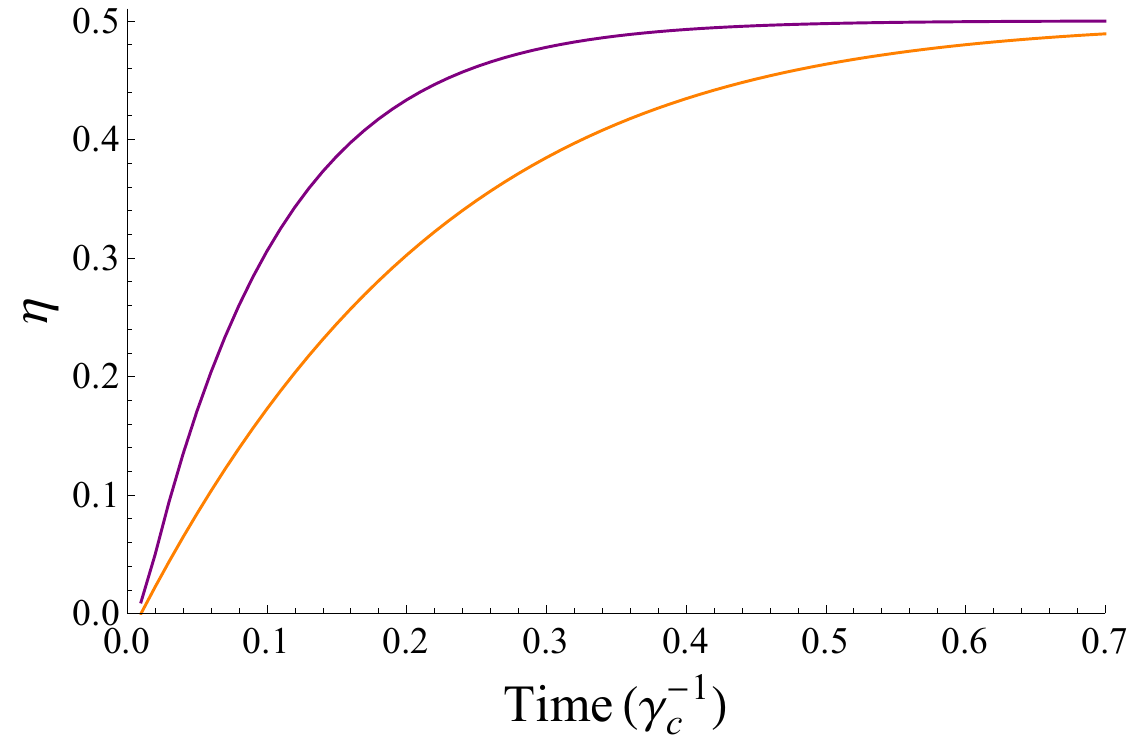}
\caption{Plot shows the value of $\eta$ as given by (\ref{eta}) versus time for a $3$-qubit dephasing example with rates $\gamma_1=\gamma_2=0.2\gamma_c$, $\gamma_3=\gamma_c$. The evolution of $\eta$ for the original dephasing code $P_3=|+++\rangle\langle +++|+|---\rangle\langle ---|$ shown in purple/above and for the new code $P'_3=|0++\rangle\langle 0++|+|1--\rangle\langle 1--|$ in orange/below.}
\label{Fig:FidBound}
\end{figure}

When the inequality (\ref{3ineq}) is satisfied the code $P'_3$ is expected to improve upon the original code.
As shown in Appendix \ref{Sec:AppFid}, the fidelity bound Eq.~(\ref{eta}) agrees in that recommendation.
The improvement of $P'_3$ as compared to $P_3$ is exemplified in Figure~\ref{Fig:FidBound}, where $\eta$ (as defined in Eq.~(\ref{eta})) is displayed as a function of time for the
specific rates $\gamma_1=\gamma_2=0.2\gamma_c$, $\gamma_3=\gamma_c$;
the corresponding time-dependent probabilities $p_i$ are given in Table \ref{probEv} in Appendix \ref{App:tPlots}.
The achieved improvement in the fidelity bound confirms that minimizing $\delta_c$ indeed permits the identification of optimal codes.

\subsection{Relationship between entanglement decay and AQEC}
\label{sec:entanglement}
From the monotonicity of entanglement, good correctability implies small decay of negativity. We now demonstrate the direct correlation between entanglement decay rate and the rate of deviation from the conditions for complete correction as given by $\delta_c$ (\ref{fn}). With the $3$-qubit dephasing encoding, the entangled state (\ref{Kappa}) becomes:
\begin{eqnarray}
|\kappa_3\rangle=\frac{1}{\sqrt{2}}(|\phi_1,+++\rangle+|\phi_2,---\rangle),\label{Bell}
\end{eqnarray}
At $t=0$ we start with the maximally entangled state $\rho=|\kappa_3\rangle\langle\kappa_3|$, for which the reduced density matrix $\rho_{red}$ for the code (obtained by tracing out the ancilla) is a projector. We thus explore the relationship between the dynamics that cause deterioration of the entanglement of $\rho$ and the behaviour of (\ref{fn}) by replacing $P$ in (\ref{PEEPe}) with $\rho_{red}$ such that
\begin{eqnarray}
\rho_{red}E_i^\dagger E_j\rho_{red}=\alpha_{ij}\rho_{red}+\Lambda_{ij}.\label{PEEPrho}
\end{eqnarray}
The resulting general expression for $\Lambda_{ij}$ is:
\begin{eqnarray}
\Lambda_{ij}=\rho_{red}E_i^\dagger E_j \rho_{red}-\frac{\mbox{Tr}(\rho_{red}E^\dagger_i E_j \rho_{red})}{\mbox{Tr}(\rho_{red})}\rho_{red}.\label{Lambdarho}
\end{eqnarray}

\begin{figure}[h!]
\centering
\includegraphics[width=10cm]{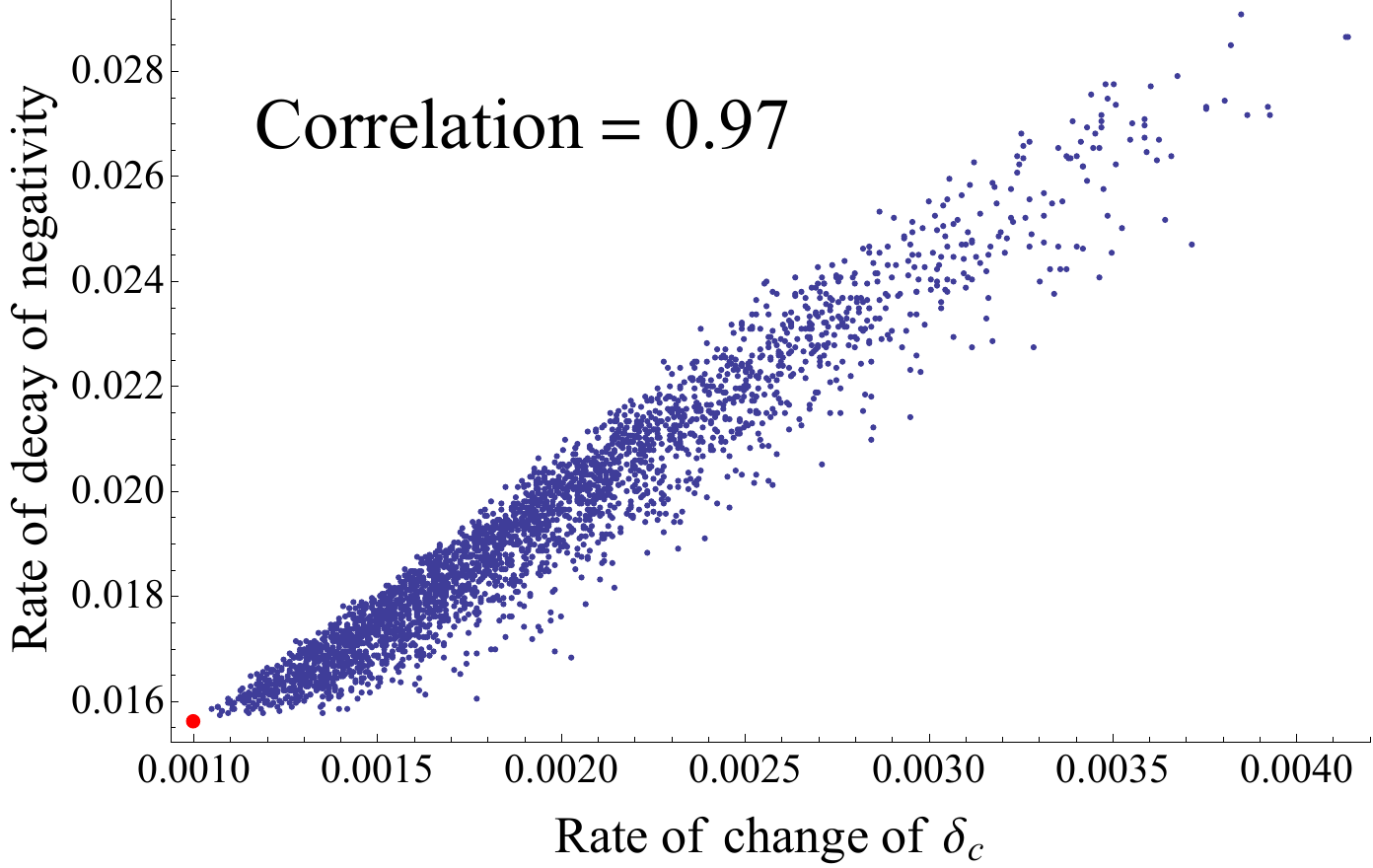}
\caption{Scatterplot for $3$-qubit dephasing channel shows strong correlation (correlation coefficient $0.97$) between the rate of deviation from (\ref{PEEP}) as given by (\ref{fn}) using (\ref{Lambdarho}), versus rate of decay of entanglement of $\rho$. Each point on the plot represents the value of the initial rate of change of the function for a different random unitary transformation of the code \cite{ZyczkowskiKus}. The point $(0.0010,0.0156)$ corresponding to the optimal choice of unitary transformation as found in-text is indicated with a larger, red dot closest to the origin, confirming the sampling over random unitaries is in the appropriate regime. ($3000$ points shown).}
\label{Fig:Correlation}
\end{figure}

By comparing the rate of deviation from (\ref{PEEP}) (as given by (\ref{fn}) using $\rho_{red}$ in (\ref{Lambdarho})), with the rate of entanglement decay for $\rho$, we find that the two have near-perfect correlation (the correlation coefficient is $0.97$ for a sample of $3000$ different unitaries). This is shown in Fig.~\ref{Fig:Correlation} for the case of a full dephasing channel (Table \ref{zchannel}) with three-qubit codes and the same particular but arbitrary choice of Lindblad rates as in Fig.~\ref{Fig:FidBound}, $\gamma_1=\gamma_2=0.2\gamma_c$ and $\gamma_3=\gamma_c$. The correlation is generated by sampling over different codes $P_n'$ obtained by random unitary transformations of the original code as in (\ref{P'}),
where the $U$ are drawn from a circular unitary ensemble \cite{ZyczkowskiKus}, and also produces the associated new input states
\begin{eqnarray}
|\kappa_n'\rangle=I\otimes U|\kappa_n\rangle.\label{kappaU}
\end{eqnarray}
While perfect correlation is not to be expected, the strength of correspondence again confirms the hypothesis that $\delta_c$ be minimal for good codes, since an increase in the rate of change of (\ref{fn}) has an associated increase in the rate of entanglement decay.

For the $3$-qubit case considered here, it is clear that the unitary transformation that enacts the change between the two optimal codes is the application of a local Hadamard on a single qubit. Note that this optimal rotation corresponds to point $(0.0010,0.0156)$ in Fig.~\ref{Fig:Correlation}, which is the point closest to the origin indicated with a larger red dot, confirming that the sampling over unitaries is over the required regime.

\begin{figure}[h!]
\centering
\includegraphics[width=10cm]{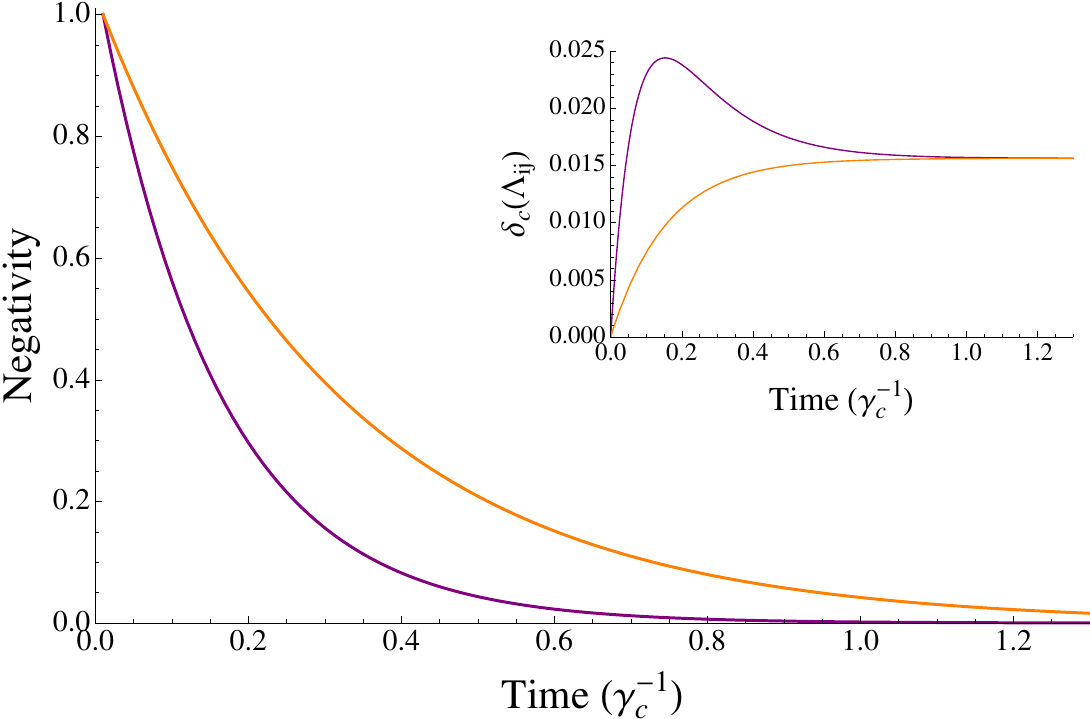}
\caption{Plot shows Negativity (\ref{Eqn:neg}) versus time for a $3$-qubit dephasing example with rates $\gamma_1=\gamma_2=0.2\gamma_c$, $\gamma_3=\gamma_c$. Entanglement decay for the original dephasing code $P_3=|+++\rangle\langle +++|+|---\rangle\langle ---|$ shown in purple/below and the decay after optimisation over unitary transformation shown in orange/above, which is coincident with the decay for the code $P'_3=|0++\rangle\langle 0++|+|1--\rangle\langle 1--|$. A significant improvement in entanglement preservation is evident.  The inset shows the corresponding evolution of the violation of the complete error correction conditions (\ref{PEEP}) as given by function (\ref{fn}) using (\ref{Lambdarho}). Here the original code is in purple/above and the optimised evolution in orange/below. }
\label{Fig:0.2_0.2_1Plots}
\end{figure}

By an exhaustive search over unitaries in (\ref{kappaU}), optimised in order to minimise entanglement decay, we find the optimisation always converges to two regimes of optimality, once again delineated by the codes identified above. In terms of the Lindblad rates, this corresponds to code $P'_3$ yielding improvement in the entanglement decay whenever rate $\gamma_3>\gamma_1,\gamma_2$ within the decay lifetime. This corresponds exactly to the regimes of inequality (\ref{3ineq}), and we provide an example plot of the entanglement decay improvement in Figure~\ref{Fig:0.2_0.2_1Plots} with the choice of rates $\gamma_1=\gamma_2=0.2\gamma_c$, $\gamma_3=\gamma_c$.

\subsection*{3.2.1. Example of recoverability}
Once optimal codes have been identified, the remaining step is to implement the appropriate syndrome measurement and recovery procedure. Consider for example the following set of operators:
\begin{eqnarray}
A_0\!&=&\!I\otimes I\otimes I,\nonumber\\
A_1\!&=&\!I\otimes \sigma_z\otimes I,\nonumber\\
A_2\!&=&\!(\sqrt{2\!-\!3q_2\!-\!q_3})I\otimes I \otimes \sigma_z - i(\sqrt{3q_2\!+\!q_3\!-\!1}) \sigma_z\otimes I\otimes \sigma_z,\nonumber\\
A_3\!&=&\!(\sqrt{1\!-\!q_3})I\otimes \sigma_z\otimes \sigma_z - i(\sqrt{q_3})\sigma_z\otimes \sigma_z\otimes \sigma_z,\hspace{0.4cm}\label{Eqn:exchan}
\end{eqnarray}
with
\begin{eqnarray}
0\leq q_2\leq \textstyle{\frac23},\hspace{0.3cm}\mbox{and}\hspace{0.3cm}0\leq q_3\leq 1, \nonumber
\end{eqnarray}
where the parameters $q_2$, $q_3$ indicate what proportion of such errors can be corrected completely. The effect of errors from this set can be completely undone with the use of the alternative code $P'_3$, that is, the conditions (\ref{PEEP}) are satisfied, which permits the identification of the correct recovery procedure in the usual way \cite{NielsenChuang}. Clearly the original code cannot satisfy (\ref{PEEP}) for this choice since it includes both two- and three-qubit errors along with the identity. We can see that if $q_3=1$, we can correct simultaneous three-qubit errors completely, while sacrificing correction on simultaneous errors on qubits two and three, with a similar relationship governing $A_2$. For the choice of operators (\ref{Eqn:exchan}) it is possible to find the set of conditions on the choice of $q_2$ and $q_3$ that will always ensure maximal recoverability of errors in the channel:
\begin{eqnarray}
p_2>p_3 \Rightarrow q_2\!&=&\!\textstyle{\frac23}, q_3=0\nonumber\\
p_2<p_3 \Rightarrow q_2\!&=&\!\textstyle{\frac13}, q_3=1\nonumber.
\end{eqnarray}
The choice (\ref{Eqn:exchan}) is thus an example where errors on multiple qubits can be corrected completely by the alternative code and not the original. 


The above method of identifying optimal codes also holds for the bitflip channel, which is unitarily equivalent to the dephasing channel. For bitflips the Lindblad equation (\ref{Lindblad}) retains terms with the identity, which implies that the evolution of the Kraus operators will depend on all four Lindblad rates $\gamma_0$, $\gamma_1$, $\gamma_2$ and $\gamma_3$. Similar improvement to the dephasing case can thus be found for bitflips, with the same relationship between the rates separating the two regimes of optimal performance: $\gamma_3>\gamma_1,\gamma_2$, i.e. the division is independent of $\gamma_0$, with the original code being $(|000\rangle\langle 000|+|111\rangle\langle 111|)$, and alternative optimal code $(|+00\rangle\langle +00|+|-11\rangle\langle -11|)$.

\subsection{Optimal $n$-qubit codes for known channels}\label{Sec:nqubit}
The generalisation to $n$-qubit repetition codes proceeds as above with appropriate extensions of the error channel and associated Lindblad rates. While in general it is desirable to keep the number of qubits to a minimum, analysis of $n$-qubit coded states gives further insight into the structures leading to optimality. Such higher-qubit systems are frequently needed in practice, and the analysis below shows that substantial improvement of the entanglement decay profile is also possible in these cases. 

As in the $3$-qubit case above, optimisation over the parameters of an initial unitary transformation of extended qubit systems reveals convergence to the same decay profile as given by two classes of optimal codes: the original repetition code
\begin{eqnarray}
P_n=|+^{\otimes n}\rangle\langle +^{\otimes n}|+ |-^{\otimes n}\rangle\langle -^{\otimes n}|,\label{origcode}
\end{eqnarray}
and the new code
\begin{eqnarray}
P_n'=|0,+^{\otimes n-1}\rangle\langle 0,+^{\otimes n-1}|+|1,-^{\otimes n-1}\rangle\langle 1,-^{\otimes n-1}|.\hspace{0.4cm}\label{newcode}
\end{eqnarray}
This latter notation indicates the coded state contains a single $|0\rangle$ in the logical code for $|+\rangle$, supplemented with $(n-1)$ qubits in $|+\rangle$, and the corresponding arrangement of a single $|1\rangle$ and $(n-1)$ qubits in $|-\rangle$ in the logical code for $|-\rangle$. When the probabilities of the different classes of errors (single, double etc.) are independent of the choice of qubits as is the case in our channels (e.g. Table \ref{zchannel}), it does not matter which qubit of the code contains the rotation. 

Similarly to equation (\ref{3ineq}), we can derive an $n$-qubit inequality which separates the regimes in which the two $n$-qubit codes (\ref{origcode}) and (\ref{newcode}) produce optimal behaviour (e.g. slowest rate of entanglement decay):
\begin{eqnarray}
\sum_{i=0}^{n}\binom{n}{i} \sqrt{p_i p_{n-i}}> 2\sum_{i=0}^{n-1}\binom{n\!-\!1}{i} \sqrt{p_i p_{i+1}},\label{genineq}
\end{eqnarray}
where $\binom{n}{i}$ are binomial coefficients. When the inequality is satisfied, the new code (\ref{newcode}) improves upon the performance of (\ref{origcode}).

\begin{figure}[h!]
\centering
\includegraphics[width=12cm]{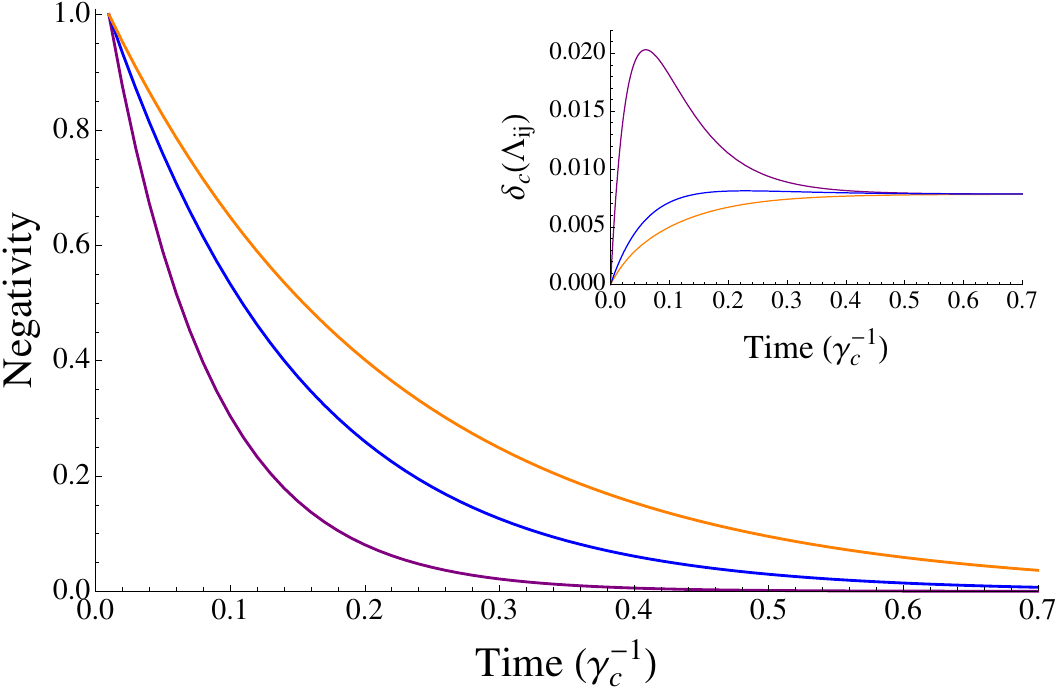}
\caption{Plot shows Negativity (\ref{Eqn:neg}) versus time for a $4$-qubit example with rates $\gamma_1=0.2\gamma_c$, $\gamma_2=0.3\gamma_c$, $\gamma_3=0.1\gamma_c$, $\gamma_4=2\gamma_c$. An indication of the evolution of the probabilities of the associated errors is given in Table \ref{4probEv}. Entanglement decay for initial codes $P_4=|+^{\otimes 4}\rangle\langle+^{\otimes 4}|+|-^{\otimes 4}\rangle\langle-^{\otimes 4}|$ shown in purple/below,  code $P_4^{''}=|+,-^{\otimes 3}\rangle\langle +,-^{\otimes 3}|+|+^{\otimes 3},-\rangle\langle +^{\otimes 3},-|$ in blue/middle, and code $P_4^{'}=|0,+^{\otimes 3}\rangle\langle 0,+^{\otimes 3}|+|1,-^{\otimes 3}\rangle\langle 1,-^{\otimes 3}|$ in orange/above. A significant improvement in entanglement preservation is evident.  The inset shows the corresponding evolution of the violation of the complete error correction conditions (\ref{PEEP}) as given by function (\ref{fn}) using (\ref{Lambdarho}) with purple as the top line, blue in the middle and orange below.}
\label{Fig:0.5_0.2_0.5_50_Plots}
\end{figure}

The utility of the method is underlined by considering an $n=4$ example in Fig.~\ref{Fig:0.5_0.2_0.5_50_Plots}, in which we compare the $4$-qubit code (\ref{origcode}) correcting single qubit errors, the code $P_4^{''}=|+,-^{\otimes 3}\rangle\langle +,-^{\otimes 3}|+|+^{\otimes 3},-\rangle\langle +^{\otimes 3},-|$ which corrects errors on four qubits, and the code (\ref{newcode}) with $4$ qubits, $P'_4$, which \textit{also} corrects correlated errors on all four, but was found in the text to be optimal also for the intermediate range. We have chosen Lindblad rates $\gamma_1=0.2\gamma_c$, $\gamma_2=0.3\gamma_c$, $\gamma_3=0.1\gamma_c$, $\gamma_4=2\gamma_c$ (the associated indicative probability evolution is given in Table \ref{4probEv} in Appendix \ref{App:tPlots}), and it is evident that the new code (\ref{newcode}) significantly outperforms both of the previous approaches.

\section{Outlook}
\label{sec:Discussion} 
The general optimisation problem of minimising $\Lambda_{ij}$ in (\ref{PEEPe}) sets the framework for finding the optimal quantum error correcting codes for protecting information where the quantum channel includes correlated errors on multiple qubits, a field still largely in its infancy. The method demonstrates significant improvement in fidelity bound and results in optimal entanglement decay profile, with recovery-independent performance evaluation of different codes and analytic results for standard examples. Such tools are invaluable in determining the choice of code in practice, where every additional qubit required for quantum coding is a significant barrier to implementation.

We have presented the explicit details for optimal performance in cases where there is a single type of error, such as dephasing. The same approach, however, also applies directly to situations in which different types of errors affect the qubits. The presence of multiple types of errors (\textit{i.e.} bitflip in addition to dephasing) typically requires additional overhead in the number of qubits that realise a code. Whereas the minimal number of qubits to define a code to protect against arbitrary single-qubit errors is $5$ \cite{Bennettetal1996, Laflammeetal1996}, practical realizations often use even more qubits such as the $7$-qubit Steane code \cite{Steane1996} or the $9$-qubit Shor code \cite{Shor1995}. Since rates for bit-flips are typically substantially lower than those for dephasing, approximate error correction certainly defines pathways to work with fewer qubits than necessary for perfect correction while improving overall performance.

For general quantum information processing, it is essential to define tools that work for arbitrary states. In more specific applications such as quantum simulations [31-33], 
it might be desirable to stress the correctability of certain states more than others. Despite the tremendous usefulness of inequality (\ref{genineq}), it has no generalization to cases where the projector $P$ in (\ref{PEEPe}) is replaced by a density matrix $\varrho$ with different weights for different components, which would permit preference to be given to certain states. In that case, optimization of reversibility via the violation of (\ref{PEEP}) is no longer possible, but defining an entangled state $\ket{\kappa}$ between code space and ancilla still enables the assessment of reversibility via entanglement decay.

\subsection*{Acknowledgements}
The authors acknowledge funding from the European Research Council within grant number 259264, and wish to thank Joonwoo Bae and Federico Levi for fruitful discussions.


\begin{appendix}
\setcounter{equation}{0}
\renewcommand{\theequation}{{A}-\arabic{equation}}

\section{Supplementary information for examples of dephasing channels}
\label{App:tPlots}
The operation elements of the full three-qubit dephasing channel are given in Table~\ref{zchannel}.
\begin{table}[h!]\centering
\begin{tabular}{l |l}
$E_0$ &$\sqrt{p_0} I\otimes I \otimes I$\\\\
$E_1$ & $\sqrt{p_1} \sigma_z \otimes I \otimes I$\\
$E_2$ & $\sqrt{p_1} I \otimes \sigma_z \otimes I$\\
$E_3$ & $\sqrt{p_1} I \otimes I\otimes \sigma_z$\\\\
$E_4$ & $\sqrt{p_2} \sigma_z \otimes \sigma_z \otimes I$\\
$E_5$ & $\sqrt{p_2} \sigma_z \otimes I \otimes \sigma_z$\\
$E_6$ & $\sqrt{p_2} I \otimes \sigma_z \otimes \sigma_z$\\\\
$E_7$ &$ \sqrt{p_3} \sigma_z \otimes \sigma_z \otimes \sigma_z$
\end{tabular}
\caption{Table of error set for full dephasing channel for three-qubit codes.}
\label{zchannel}
\end{table}

Explicit expressions for the time-dependent probabilities of errors occurring in noisy channels can be found by equating the Lindblad form with the Bloch representation of $\rho$ in the standard way (see e.g. \cite{NielsenChuang}). For the full dephasing channel given in Table \ref{zchannel}, the explicit expressions for the probabilities in terms of the Lindblad rates are given in (\ref{3ps}). 

Indicative values of the evolution of probabilities (\ref{3ps}) for a particular choice of Lindblad rates where improvements can be made to the standard dephasing code are given in Table~\ref{probEv}. Similarly, indicative values of the probability evolution for $n=4$ with Lindblad rates $\gamma_1=0.2\gamma_c$, $\gamma_2=0.3\gamma_c$, $\gamma_3=0.1\gamma_c$, $\gamma_4=2\gamma_c$ are given in Table~\ref{4probEv}.
\begin{eqnarray}
\sqrt{p_0}&=& \frac{1}{2\sqrt{2}}\Big(e^{-4 (3 \gamma_1+2 \gamma_{2}+\gamma_{3}) t}\Big.\nonumber\\
&&\left.\times \sqrt{3 e^{8 (2 \gamma_{1}+\gamma_{2}+\gamma_{3}) t}+e^{8 (3 \gamma_{1}+2 \gamma_{2}+\gamma_{3}) t}+3 e^{4 (5 \gamma_{1}+2 \gamma_{2}+\gamma_{3}) t}+e^{4 (3 \gamma_{1}+4 \gamma_{2}+\gamma_{3}) t}}\right),\nonumber\\
\sqrt{p_1}&=&  \frac{1}{2\sqrt{2}}\left(\sqrt{1-e^{-8 (\gamma_{1}+\gamma_{2}) t}-e^{-4 (3 \gamma_{1}+\gamma_{3}) t}+e^{-4 (\gamma_{1}+2 \gamma_{2}+\gamma_{3}) t}}\right),\nonumber\\
\sqrt{p_2}&=& \frac{1}{2\sqrt{2}}\left(\sqrt{1-e^{-8 (\gamma_{1}+\gamma_{2}) t}+e^{-4 (3 \gamma_{1}+\gamma_{3}) t}-e^{-4 (\gamma_{1}+2 \gamma_{2}+\gamma_{3}) t}}\right),\nonumber\\
\sqrt{p_3}&=&  \frac{1}{2\sqrt{2}}\left(\sqrt{1+3 e^{-8 (\gamma_{1}+\gamma_{2}) t}-e^{-4 (3 \gamma_{1}+\gamma_{3}) t}-3 e^{-4 (\gamma_{1}+2 \gamma_{2}+\gamma_{3}) t}}\right).\label{3ps}
\end{eqnarray}

\begin{table}[h!]\centering
\begin{tabular}{l |c|c|c|c|c}
&$t=0$&$t=0.1$&$t=0.2$&$t=0.4$&$t=0.6$\\
\hline
$p_0$&$1$&$0.66$&$0.46$&$0.28$&$0.19$\\
$3p_1(\equiv 3p_2)$&$0$&$0.10$&$0.18$&$0.27$&$0.32$\\
$p_3$&$0$&$0.13$&$0.18$&$0.19$&$0.17$
\end{tabular}
\caption{Indicative approximate values of the probability evolution of (\ref{3ps}) for a $3$-qubit repetition code sent through a full dephasing channel (Table \ref{zchannel}) with rates $\gamma_1=\gamma_2=0.2\gamma_c$, $\gamma_3=\gamma_c$ as in Figs.~\ref{Fig:FidBound} \& \ref{Fig:0.2_0.2_1Plots}. Here the probabilities of single, double and triple errors are approximately equal for short times, and single and double errors begin to dominate at long times although the contribution from $p_3$ remains significant.}
\label{probEv}
\end{table}
\begin{table}[h!]\centering
\begin{tabular}{l |c|c|c|c|c|c}
&$t=0$&$t=0.05$&$t=0.1$&$t=0.2$&$t=0.3$&$t=0.4$\\
\hline
$p_0$&$1$&$0.62$&$0.41$&$0.21$&$0.13$&$0.09$\\
$4p_1$&$0$&$0.06$&$0.11$&$0.16$&$0.19$&$0.21$\\
$6p_2$&$0$&$0.15$&$0.24$&$0.34$&$0.38$&$0.39$\\
$4p_3$&$0$&$0.04$&$0.09$&$0.15$&$0.19$&$0.21$\\
$p_4$&$0$&$0.12$&$0.16$&$0.14$&$0.11$&$0.09$
\end{tabular}
\caption{Indicative approximate values of the probability evolution according to a $4$-qubit example of a full dephasing channel with rates $\gamma_1=0.2\gamma_c$, $\gamma_2=0.3\gamma_c$, $\gamma_3=0.1\gamma_c$, $\gamma_4=2\gamma_c$ corresponding to Fig.~\ref{Fig:0.5_0.2_0.5_50_Plots}.}
\label{4probEv}
\end{table}

\section{Comparison of fidelity bound and inequality~(\ref{3ineq})}
\label{Sec:AppFid}
Here we show that the fidelity bound as given by (\ref{eta}) and the inequality~(\ref{3ineq}) agree in their assessment of the codes for $3$-qubit dephasing channels. 

The assessment of (\ref{eta}) requires the maximization over input states and we found the states $(\ket{+^{\otimes 3}}+i\ket{-^{\otimes 3}})/4$ to achieve the maximum both for code $P_3$ and $P_3^\prime$ defined in Eqs.~\ref{dephcode} and \ref{newcode1}.


With these states, one readily obtains the bounds
\begin{eqnarray}
\eta_{P_3}=\frac{6p_1p_2}{p_1+p_2}+\frac{2p_0p_3}{p_0+p_3}
\end{eqnarray}
and
\begin{eqnarray}
\eta_{P_3^\prime}=\frac{2p_0p_1}{p_0+p_1}+\frac{4p_1p_2}{p_1+p_2}+\frac{2p_2p_3}{p_2+p_3}.
\end{eqnarray}
The fidelity bounds thus predict the code $P_3^\prime$ to be better than $P_3$ if
\begin{eqnarray}
\eta_{P_3}-\eta_{P_3^\prime}=2\left(\frac{p_0p_3}{p_0+p_3}+\frac{p_1p_2}{p_1+p_2}-\frac{p_0p_1}{p_0+p_1}-\frac{p_2p_3}{p_2+p_3}\right) .
\end{eqnarray}
is postive.

On the other hand, inequality~(\ref{3ineq}) predicts $P_3^\prime$ to be better than $P_3$ if
\begin{eqnarray}
\mathcal{X}=\sqrt{p_0p_3}+\sqrt{p_1p_2}-\sqrt{p_0p_1}-\sqrt{p_2p_3}
\end{eqnarray}
is positive.

Multiplying $(\eta_{P_3}-\eta_{P_3^\prime})$ and $\mathcal{X}$ one obtains
\begin{eqnarray}
\hspace{-2.6cm}(\eta_{P_3}-\eta_{P_3^\prime})\mathcal{X}=2\frac{p_0p_2p_3+p_0p_1p_2+p_0p_1p_3+p_1p_2p_3}{(p_0+p_1)(p_0+p_3)(p_1+p_2)(p_2+p_3)}(p_0\!-\!p_2)(\sqrt{p_0}\!-\!\sqrt{p_2})(p_1\!-\!p_3)(\sqrt{p_1}\!-\!\sqrt{p_3}) ,\nonumber\\
\end{eqnarray}
which is non-negative since $p_i\ge 0$.
That is, there is no case where $\mathcal{X}$ is negative and $(\eta_{P_3}-\eta_{P_3^\prime})$ positive, \textit{i.e.}, there is no case where the two conditions would recommend a different code.

\end{appendix}

\subsection*{References}

\bibliographystyle{unsrt}

\begin{thebibliography}{widest-label}
\footnotesize

\bibitem{BlochReview}
I. Bloch, J. Dalibard and W. Zwerger.
\newblock {\em Rev. Mod. Phys.}, \textbf{80}, 885 (2008).

\bibitem{MintertWunderlich}
F. Mintert and C. Wunderlich.
\newblock {\em Phys. Rev. Lett.}, \textbf{87}, 257904 (2001).

\bibitem{Aliferisetal2006}
P. Aliferis, D. Gottesman and J. Preskill.
\newblock {\em Quant. Inf. Comput.}, \textbf{6}, 97, (2006).

\bibitem{Terhal2005}
B.M. Terhal and G. Burkard.
\newblock {\em Phys. Rev. A}, \textbf{71}, 012336 (2005).

\bibitem{KlesseFrank2005}
R. Klesse and S. Frank.
\newblock {\em Phys. Rev. Lett.}, \textbf{95}, 230503 (2005).

\bibitem{Aharonovetal2006}
D. Aharonov, A. Kitaev and J. Preskill.
\newblock {\em Phys. Rev. Lett.}, \textbf{96}, 050504 (2006).

\bibitem{Bombinetal2012}
H. Bombin et al.
\newblock {\em Phys. Rev. X}, \textbf{2}, 021004 (2012).

\bibitem{Leungetal1997}
D.W. Leung et al.
\newblock {\em Phys. Rev. A}, \textbf{56}, 2567 (1997).

\bibitem{FTQEC}
A.Y. Kitaev.
\newblock {\em in Quantum Communication, Computing and Measurement}, O. Hirota et. al. Eds., Plenum, New York (1997);
E. Knill, R. Laflamme and W.H. Zurek.
\newblock {\em Science}, \textbf{279}, 342 (1998);
D. Aharonov and M. Ben-Or. 
\newblock {\em SIAM J. Comput.}, \textbf{38}, 1207 (2008).

\bibitem{KnillLaflamme}
E. Knill and R. Laflamme.
\newblock {\em Phys. Rev. A}, \textbf{55}, 900 (1997).

\bibitem{NielsenChuang}
M.A. Nielsen and I.L. Chuang.
\newblock Quantum Computation and Quantum Information.
\newblock {\em Cambridge: Cambridge University Press}, 2010.

\bibitem{NgMandayam}
H.K. Ng and P. Mandayam.
\newblock {\em Phys. Rev. A}, \textbf{81}, 062342 (2010).

\bibitem{MandayamNg}
P. Mandayam and H.K. Ng.
\newblock {\em Phys. Rev. A}, \textbf{86}, 012335 (2012).

\bibitem{SW2002}
B. Schumacher and M.D. Westmoreland.
\newblock {\em Quant. Inf. Proc.}, \textbf{1}, 5 (2002).

\bibitem{BarnumKnill2002}
H. Barnum and E. Knill.
\newblock {\em J. Math. Phys.}, \textbf{43}, 2097 (2002).

\bibitem{Taylor2010}
J. Taylor.
\newblock {\em J. Math. Phys.}, \textbf{51}, 092204 (2010).

\bibitem{BenyOreshkov2010}
C. B\'eny and O. Oreshkov.
\newblock {\em Phys. Rev. Lett.}, \textbf{104}, 120501 (2010).

\bibitem{WickertANDvLoock}
R. Wickert, N.K. Bernardes and P. van Loock.
\newblock {\em Physi. Rev. A}, \textbf{81}, 062344 (2010).

\bibitem{Muraoetal}
M. Murao, M.B. Plenio and V. Vedral.
\newblock {\em Phys. Rev. A}, \textbf{61}, 032311 (2000).

\bibitem{SainzBjork}
I. Sainz and G. Bj\"ork.
\newblock {\em Phys. Rev. A}, \textbf{77}, 052307 (2008).

\bibitem{HorodeckiReview}
R. Horodecki et al.
\newblock {\em Rev. Mod. Phys.}, \textbf{81}, 865 (2009).

\bibitem{Schumacher1996}
B. Schumacher.
\newblock {\em Phys. Rev. A}, \textbf{54}, 2614 (1996).

\bibitem{VidalWerner}
G. Vidal and R.F. Werner.
\newblock {\em Phys. Rev. A}, \textbf{65}, 032314 (2002).

\bibitem{Shor1995}
P.W. Shor.
\newblock {\em Phys. Rev. A}, \textbf{52}, R2493 (1995).

\bibitem{Reedetal2012}
M.D. Reed et al.
\newblock {\em Nature}, \textbf{482}, 382 (2012).

\bibitem{AolitaAcin2010}
L. Aolita et al.
\newblock {\em Phys. Rev. A}, \textbf{82}, 032317 (2010).

\bibitem{ZyczkowskiKus}
K. Zyczkowski and M. Ku\'s.
\newblock {\em J. Phys A: Math. Gen.}, \textbf{27}, 4235 (1994).

\bibitem{Bennettetal1996}
C.H. Bennett et al.
\newblock {\em Phys. Rev. A}, \textbf{54}, 3824 (1996).

\bibitem{Laflammeetal1996}
R. Laflamme et al. 
\newblock {\em Phys. Rev. Lett.}, \textbf{77}, 198 (1996).

\bibitem{Steane1996}
A.M. Steane.
\newblock {\em Phys. Rev. Lett.}, \textbf{77}, 793 (1996).

\bibitem{BlochetalReview2012}
I. Bloch, J. Dalibard and S. Nascimb\`ene.
\newblock {\em Nat. Phys.}, \textbf{8}, 267 (2012).

\bibitem{BlattRoos2012}
R. Blatt and C.F. Roos.
\newblock {\em Nat. Phys.}, \textbf{8}, 277 (2012).

\bibitem{AspuruGandWalther2012}
A. Aspuru-Guzik and P. Walther.
\newblock {\em Nat. Phys.}, \textbf{8}, 285 (2012).





\end{thebibliography}

\end{document}